\documentclass{aa}  
\usepackage{graphicx}
\usepackage{xcolor}

\usepackage{txfonts}

\newcommand{\nc}{\newcommand}
\nc{\lsun}{\ensuremath{\mathrm{L}_\odot}}
\nc{\msun}{\ensuremath{\mathrm{M}_\odot}}
\nc{\tex}{\ensuremath{\mathrm{T}_{\rm ex}}}
\nc{\cthree}{C$_3$}
\nc{\cthreehtwo}{$c$-C$_3$H$_2$}
\nc{\kms}{\mbox{km\,s$^{-1}$}}
\nc{\Kkms}{\mbox{K\,km\,s$^{-1}$}}
\nc\micron{\mbox{$\mu$m}}
\nc{\Trot}{$T_{\rm rot}$}%
\nc{\Ntot}{$N(C_3)$}%
\nc{\Tc}{$T_{\rm c}$}%
\nc{\Tdust}{$T_{\rm dust}$}%
\nc{\Tex}{$T_{\rm ex}$}%
\nc{\Tkin}{$T_{\rm kin}$}%
\nc{\Tmax}{$T_{\rm max}$}%
\nc{\cmcub}{\mbox{cm$^{-3}$}}
\nc{\cmsq}{\mbox{cm$^{-2}$}}

\newcommand{\HI}{H {\sc i}}

\newcommand\arcdeg{\mbox{$^\circ$}}%

\begin{document}

\title{Deuterium fractionation of a distant cold dark cloud along the line
  of sight of W51}

\author{C. Vastel \inst{1,2} \and B. Mookerjea\inst{3} \and J. Pety\inst{4}
  \and M. Gerin\inst{5,6}}

\institute{Universit\'e de Toulouse, UPS-OMP, IRAP, Toulouse, France
  \email{charlotte.vastel@irap.omp.eu} %
  \and CNRS, IRAP, 9 Av. Colonel Roche, BP 44346, F-31028 Toulouse Cedex 4,
  France %
  \and Tata Institute of Fundamental Research, Homi Bhabha Road, Mumbai
  400005, India %
  \and Institut de Radioastronomie Millim\'etrique, 300 Rue de la Piscine,
  F-38406 Saint Martin d'H\`eres, France %
  \and LERMA, Observatoire de Paris, PSL Research University, CNRS,
  UMR8112, Paris, France, F-75014 %
  \and Sorbonne Universit\'es, UPMC Univ. Paris 06, UMR8112, LERMA, Paris,
  France, F-75005 %
}


 
\abstract{Herschel/HIFI observations toward the compact HII region W51
  has revealed the presence of a cold dense core along its line of sight in a 
   high-velocity stream located just in front of W51. This detection
  has been made possible through absorption measurements of low-energy
  transitions of HDO, NH$_3$, and C$_3$ against the bright background
  emitted by the star-forming region. We present a follow-up study of this
  core using the high sensitivity and high spectral resolution provided by
  the IRAM 30-meter telescope. We report new detections of this core in
  absorption for DCO$^+$ (2--1, 3--2), H$^{13}$CO$^+$ (1--0), DNC (3--2),
  HN$^{13}$C (1--0), p-H$_2$CO (2$_{0,2}$--1$_{0,1}$, 3$_{0,3}$--2$_{0,2}$), and in emission
  for o-NH$_2$D. We also report interferometric observation of this
    last species using the IRAM/NOEMA telescope, revealing the fragmented nature of
    the source through the detection of two cores, separated by
    $0.19-0.24$ pc, with average sizes of less than $0.16-0.19$ pc.
  From a non-LTE analysis, we are able to estimate the density ($\sim$
  2.5~$\times$ 10$^4$ cm$^{-3}$) and temperature ($\sim$ 10 K) of this
  component, typical of what is found in dark clouds. This component
  (called W51-core) has the same DCO$^+$/HCO$^+$ ratio (0.02) as TMC-1
  and a high DNC/HNC ratio (0.14). Detection of these deuterated species
  indicates that W51-core is similar to an early-phase low-mass star-forming
  region, formed from  the interaction between the W51 giant molecular
  cloud and the high-velocity stream in front of it. The W51 complex being
  at about 5 kpc, these findings lead to what is the first detection
  of the earliest phase of low-mass star-forming region at such a large
  distance.}


\maketitle{} %


\section{Introduction}

The unprecedented sensitivity and large-scale mapping capabilities of
Herschel/SPIRE have revealed a large network of parsec-scale filaments in
Galactic molecular clouds, which indicate an intimate connection between
the filamentary structure of the ISM and the formation of dense cloud cores
\citep{andre2010,molinari2010}. Since star formation occurs mostly in
prominent filaments, characterizing the physical properties of these
regions is the key to understanding the process of star formation.

We have recently detected a dense core for the first time along the line of sight 
of the compact HII region W51 \citep{mookerjea2014}.
Having no information on the spatial extent of this
core, we could not distinguish between a core (0.03-0.2 pc) or a clump
(0.3--3 pc). Located at a distance of 5.41$^{+0.31}_{-0.28}$\,kpc
\citep{sato2010}, W51 is a radio source with a complicated morphology (see
\citealt{sollins2004}  for a sketch of the spatial distribution of the various
components described below) in which many compact sources are superimposed
on extended diffuse emission \citep{bieging1975}. The line of sight to W51
intersects the Sagittarius spiral arm nearly tangentially (l = 49\arcdeg),
which means that sources over a $\sim 5$\,kpc range of distances are
superimposed along the line of sight. \citet{carpenter1998} divided the
molecular gas associated with the W51 HII region into two subgroups: a
giant molecular cloud (1.2$\times 10^6$ M$_{\odot}$) at V$_{LSR}$ $\sim$
61~\kms, and an elongated (22$\times$136\,pc) molecular cloud similar  to a
filamentary structure (1.9$\times 10^5$ M$_{\odot}$) at 68~\kms.
\citet{koo1997} used HI observations toward W51e2 to identify two sets of
absorption features, the local features at 6.2, 11.8, and 23.1~\kms\ and
features thought to  belong to the Sagittarius spiral arm at 51.1,
62.3, and 68.8~\kms.  The massive star formation activity in the W51 region
is believed to have resulted from a collision between the W51 giant
molecular cloud and the high-velocity (HV; 68~\kms) cloud
\citep{carpenter1998,kang2010}. Using H$_2$CO observations,
\citet{arnal1985} detected five absorption components at $\sim$ 52, 55, 57,
66, and 70~\kms\ toward W51e2.  The presence of H$_2$CO absorption between
66 and 70~\kms\ and of \HI\ absorption at 68.8~\kms\ implies that the HV
stream is located in front of W51e.


The dense core was detected as an absorption feature at $\sim 69$\,\kms\
toward the compact HII region W51e2 in transitions of HDO, CN, NH$_3$, CCH,
$c$-C$_3$H$_2$, and \cthree \citep{mookerjea2014}. This was the first
ever detection of HDO in an environment outside low- and high-mass star-forming regions. Among the species detected in the dense core, HDO, $o$-
and $p$-NH$_3$, and \cthree\ show line widths between 1.5--2\,\kms, while
the lines of CN and CCH are narrower (0.6--0.7\,\kms). Non-LTE RADEX
modeling of CN and NH$_3$ also indicate the presence of two components
contributing to the absorption: NH$_3$ arises from the denser
(5\,10$^5$\,\cmcub) and warmer (30\,K) gas, while CN arises from a less
dense ($10^5$\,\cmcub) and cooler medium (6--10\,K). We thus concluded that
the 69\,\kms\ feature arises in a dense and cold core that is formed within
the much larger scale filament, deemed to be interacting with the W51 main
molecular cloud. A possible scenario for formation of the dense core was
suggested based on the collision of the filament with W51 \citep{kang2010}. In this scenario, cloud-cloud collisions lead to the
compression of the interface region and initiate the formation of stars,
causing the molecular cores at the interface to be heated, but leaving the cores on
the trailing side  cold \citep{habe_ohta1992}.

Based on the available observations, \citet{mookerjea2014} ascertained the
abundances of all detected species in the 69\,\kms\ absorption feature.
The abundances are similar to those found in the prestellar and
protostellar cores. The D/H ratio for water vapor is 9.6 $\times 10^{-4}$,
similar to the value found in Orion KL \citep{neill2013} and also similar
to the value found in the cold envelope of the IRAS 16293-2422 low-mass
star-forming region \citep{coutens2012,coutens2013} where the ices are
not affected by thermal desorption or photo-desorption by the FUV field.

In this paper we present observations from the IRAM NOEMA and 30m
  telescopes of several tracers (focused on deuterium chemistry)
toward the newly detected cold dense core to uncover the conditions
prevailing in this filament and also to probe the degree of deuteration in
the core.

\section{Observations}

\subsection{IRAM 30-meter observations}

\begin{figure}
  \centering
  \includegraphics[width=9cm]{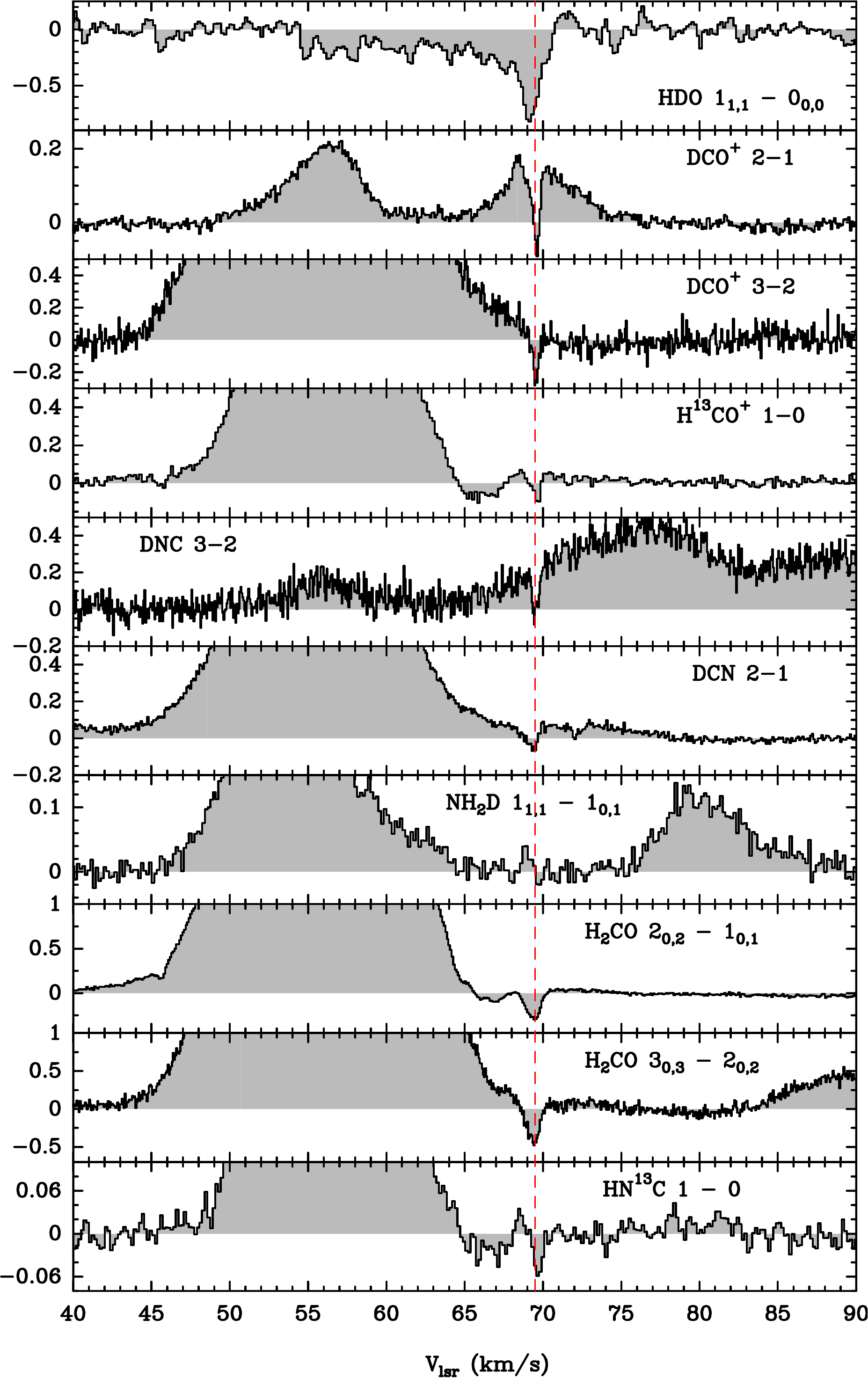}
  \caption{Herschel/HIFI (HDO) and IRAM observations (in T$_A$$^*$). The red vertical dashed line indicates the location of the core at a velocity of 69.5 km~s$^{-1}$.}
  \label{hocop_lines}
\end{figure}

The observations were performed at the IRAM 30m telescope toward W51
($\alpha_{2000} = 19^h23^m43.9^s, \delta_{2000} =
14\degr30\arcmin30.5\arcsec$) between July 16, 2013, and July 19, 2013.
We used the broadband receiver EMIR connected to an FTS spectrometer in
its 50 kHz resolution mode. The beam of the observations is 29$''$, 23$''$,
and 11$''$ at 86, 106, and 210 GHz, respectively.  Weather conditions were
average with 2 to 3 mm of precipitable water vapor. In order to obtain a
flat baseline, observations were carried out using a nutating secondary
mirror with a throw of 3 arcmin. No
contamination from the reference position was observed. \\
Pointing was checked every 1.5 hours on the nearby continuum sources
1923+210, 1749+096, and 1757-240. We adopted the telescope and receiver
parameters (main-beam efficiency, half power beam width, forward
efficiency) from the values
monitored at IRAM\footnote{http://www.iram.fr}. \\
Figure \ref{hocop_lines} presents the IRAM 30m observations of all the
  species unambiguously detected in our observations along with the
  Herschel/HIFI HDO detection from \citet{mookerjea2014}. A red dashed line indicates the location of the core at 69.5 km~s$^{-1}$. Line intensities
are expressed in units of antenna temperatures, before correction for the
main beam efficiencies.  The continuum has been subtracted and its values
are quoted in Table \ref{lines}.  Table \ref{lines} lists the spectroscopic
parameters of the detected transitions, the spectral resolution
($\Delta$V), as well as the properties of the observed transitions based
on Gaussian fitting of the lines: V$_{center}$ (velocity centroid of the
Gaussian fitting function), T$_C$ (continuum value), FWHM (full width at
half maximum).  The integrated opacity for the absorption lines is
calculated in column 9 and is computed as
\begin{equation}
  \int\tau_{obs}dV = -\int ln\frac{T_{mb}}{T_{C}}dV
.\end{equation}

If the line is in emission, column 9 simply states the integrated area in
K~km~s$^{-1}$, assuming that the foreground core responsible for the
  absorption of the continuum is more extended than the continuum emitting
  region. Indeed, as long as the size of the absorbing layer is larger than
  the size of the region emitting the continuum, the line-to-continuum
  ratio does not depend on the sizes of the telescope beam used for the
  observations.  We used the CASSIS\footnote{http://cassis.irap.omp.eu}
software \citep{vastel2015}, developed at IRAP, for the line
identification.

\begin{table*} 
  \small
  \caption{Properties of the observed transitions at the IRAM 30-meter telescope. The spectroscopic parameters are from CDMS. Properties of the observed transitions at the IRAM 30-meter telescope. The spectroscopic parameters are from CDMS. Data are available at the CDS via http://cdsweb.u-strasbg.fr/cgi-bin/qcat?J/A+A/.\label{lines}}
  \begin{tabular}{cccccccccccc}
    Species   &Transitions  &  Frequency & $\Delta V$  & E$_{low}$ & E$_{up}$  &  V$_{center}$       &  T$_C$    &   FWHM               &  $\int \tau_{obs}dV$ or area & rms  \\
    &                   &     (MHz)      &  (km~s$^{-1}$)  & (K) & (K)          &  (km~s$^{-1}$)     &    K          &(km~s$^{-1}$)       &     (km~s$^{-1}$) or (K~km~s$^{-1}$) & (mK) \\
    \hline
    \hline
    DCO$^+$   &   2--1  & 144077.2890  & 0.10  &3.46 & 10.37  & 69.58 $\pm$ 0.02   & 1.08   & 0.40 $\pm$ 0.05  &  0.17 $\pm$ 0.03 & 14.8 \\
    DCO$^+$   &   3--2  & 216112.5822  & 0.07  &10.37 & 20.74  & 69.56 $\pm$ 0.03   & 2.58   & 0.36 $\pm$ 0.06  &  0.12 $\pm$ 0.02 &89.8 \\
    H$^{13}$CO$^+$  &   1--0  & 86754.2884  & 0.17 & 0.00 & 4.16  & 69.64 $\pm$ 0.10   & 0.93   & 0.34 $\pm$ 0.11  &  0.11 $\pm$ 0.02  & 20.0\\
    DNC  &   3--2  & 228910.4810  & 0.07    & 11.00 & 21.97  & 69.52 $\pm$ 0.03   & 2.54   & 0.42 $\pm$ 0.06  &  0.09 $\pm$ 0.01  &99.9 \\
    HN$^{13}$C  & 1--0  &  87090.8252  &  0.17 & 0.00 & 4.18  & 69.70 $\pm$ 0.03  & 0.93 & 0.54 $\pm$ 0.07  & 0.010 $\pm$ 0.002  &16.2  \\
    o-NH$_2$D  &   1$_{1,1}$--1$_{0,1}$  & 85926.2780  & 0.17   & 16.56 & 20.68  & 68.91 $\pm$ 0.06   & 1.20   & 0.58 $\pm$ 0.15  &  0.020 $\pm$ 0.003\tablefootmark{a} &15.7 \\
    p-H$_2$CO  &   2$_{0,2}$--1$_{0,1}$  & 145602.9490  & 0.10  & 3.50 &10.48  & 69.38 $\pm$ 0.02   & 1.08   & 0.93 $\pm$ 0.04  &  0.50 $\pm$ 0.08 &24.9 \\
    p-H$_2$CO  &   3$_{0,3}$--2$_{0,2}$  & 218222.1920  & 0.07  & 10.48 & 20.96  & 69.39 $\pm$ 0.03   & 2.42   & 0.78 $\pm$ 0.06  &  0.39 $\pm$ 0.06 &78.4 \\
    \hline
  \end{tabular}
  \tablefoot{\tablefoottext{a}{Only transition detected in emission}}
\end{table*}

\subsection{IRAM NOEMA observations}

 To resolve the spatial distribution of the source that shows narrow 
  absorption of deuterated species, we obtained additional observations of
  the o-NH$_2$D emission from the NOEMA interferometer in C and D
  configurations. We obtained 10 hours of usable on-source data (time
  computed for an eight-antenna array). Weather varied from excellent to
  average. The achieved resolution is $5.36 \times 4.77''$ at a position
  angle of $-20.4\arcdeg$. The lower sideband of the 3\,mm receivers was
  tuned at 85.926 GHz. One of the narrowband correlator windows was
  centered at the frequency of the o-NH$_2$D line (85.926\,GHz) yielding
  20\,MHz-wide spectra at 39\,kHz (or 0.14\,km\,s$^{-1}$) channel
  spacing. We observed a single field centered on the M51 HII compact
  region. The NOEMA primary beam size is typically $59''$ at the observing
  frequency.

  MWC\,349 was used as flux calibrator, 1920$+$154 and 1932$+$204 as
  phase/amplitude calibrators, and 3C454.3 as bandpass calibrator.  Data
  calibration and imaging/deconvolution used the standard procedures of the
  \textsc{CLIC} and \textsc{MAPPING} software of the \textsc{GILDAS}
  suite\footnote{See \texttt{http://www.iram.fr/IRAMFR/GILDAS/} for more
    details on the \textsc{GILDAS} software.}. Data was self-calibrated in
  phase on the strong continuum source toward W51. No short-spacing
  corrections were applied.

\begin{figure}
  \centering
  \includegraphics[width=9cm]{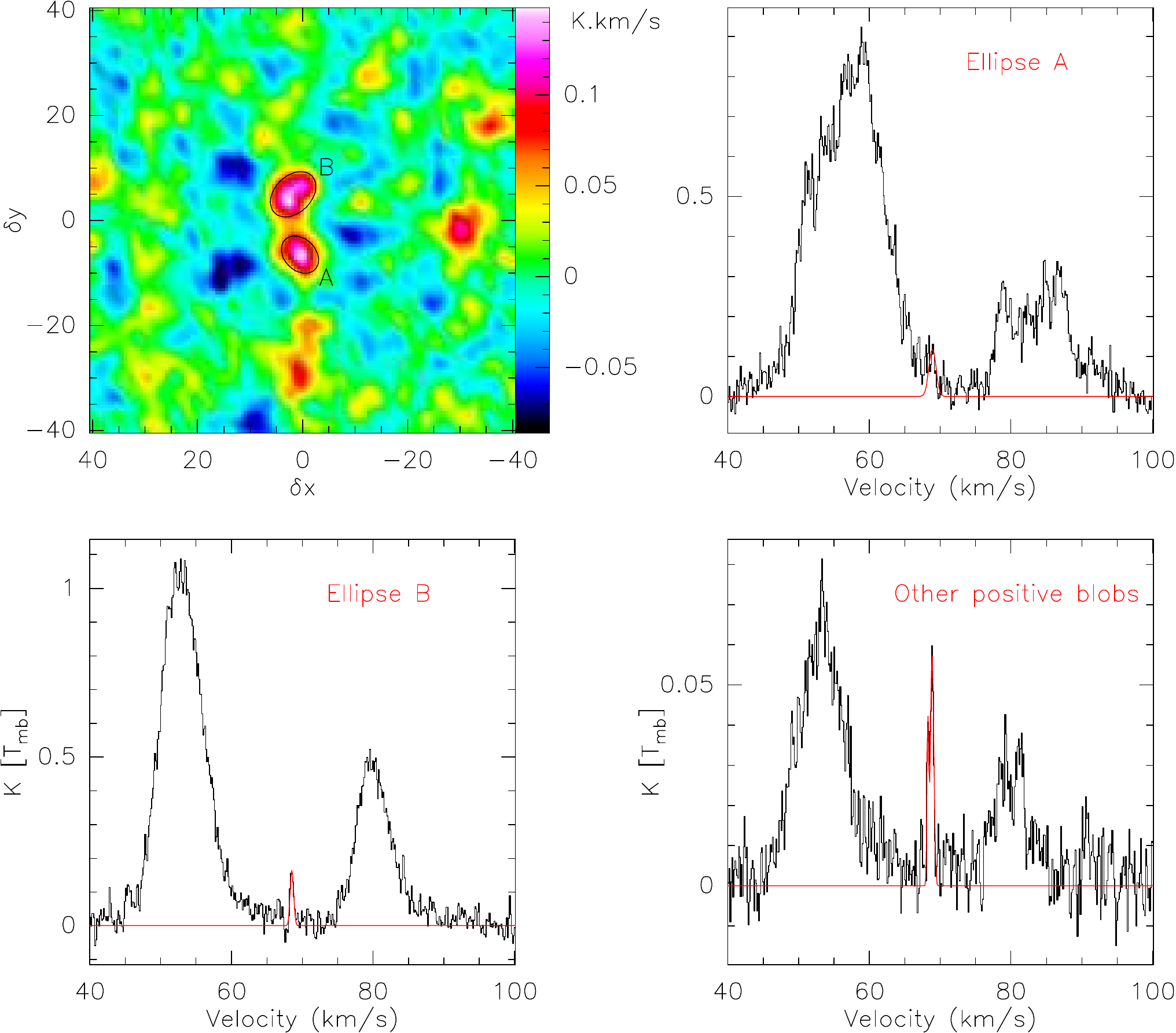}
  \caption{ Results from the NOEMA observations for the NH$_2$D
      transition at 86 GHz. Top left: Line intensity integrated between
      68.1 and 69.0 km~s$^{-1}$. Top right: Spectra integrated over ellipse
      A. Bottom left: Spectra integrated over ellipse B. Bottom right:
      Spectra integrated over all pixels of the top left image whose
      integrated intensity belongs to the [0.025, 0.2 K~km~s$^{-1}$] range,
      except pixels inside ellipse A and B.}
  \label{nh2d}
\end{figure}

   As shown in Fig.~\ref{nh2d}, the NH$_2$D emission integrated between 
   68.1 and 69.0 km\,s$^{-1}$ is fragmented into several peaks. 
   Two marginally resolved sources (A and B) are clearly
  detected. About 15\% of the total flux is scattered outside these two
  main sources. The NH$_2$D line sits  between two wide lines, identified 
  as CH$_3$OCHO (7(6,2)--6(6,1)) at 85.927 GHz and CH$_3$OCHO (7(6,1)--6(6,0)) at 85.919
  GHz. Both lines are emitted at the velocity of the W51 HII
  region ($\sim$ 55 km~s$^{-1}$).

  Table~\ref{pdb} presents the observational properties of both fragments,
  integrated over two ellipses  $10\times7''$ and $8\times6''$ in size for A
  and B, respectively. The IRAM 30m beam size is $30.2''$. This gives beam
  dilution factors of 13 and 19, respectively. Applying these factors to
  the measured integrated intensities, we find that we recover
  17\,mK\,km\,$^{-1}$, i.e., the intensity measured at the resolution of
  the IRAM 30m, within the uncertainties.

\begin{table} 
  \caption{Properties of the observed NH$_2$D transition at the IRAM
      Plateau de Bure interferometer. \label{pdb}}
  \begin{tabular}{ccccc}
    Location   &    V$_{center}$    &  Area                   &        FWHM               &  T$_{peak}$   \\
    &     (km~s$^{-1}$)  &  (mK.km~s$^{-1}$)    &  (km~s$^{-1}$)         &        (mK)            \\
    \hline
    \hline
    Ellipse A  & 68.90  & 150.2 & 1.20 & 117.58   \\
    Ellipse B  & 68.54 & 100.3 & 0.57 & 164.45 \\
    Other       & 68.25 &  13.1  & 0.30 & 40.91   \\
    & 68.85 &  33.8  & 0.55 & 57.71 \\
    \hline
  \end{tabular}
\end{table}

\section{Discussion}

\subsection{Radiative transfer modeling}

From \citet{mookerjea2014}, we were able to identify two physical
components: one warm component (30 K)  with a H$_2$ density of 5~10$^{5}$
cm$^{-3}$ detected through NH$_3$ measurements, while the other component
is less dense (10$^{5}$ cm$^{-3}$) with a lower kinetic temperature (6 K),
detected with CN. However, owing to the complexity of the line profiles
  (see Figs. 3 and 4 from Mookerjea et al. 2014), which were explained by the
  contamination by the molecular cloud complex between 55 and 70 km/s, the
  H$_2$ density and kinetic temperature estimated from non-LTE modeling
  have a large error bar (valid to within 50\%).

For the following analysis we used the CASSIS software, which performs LTE
and non-LTE modeling and list in Table \ref{radtrans} the results from our
radiative transfer analysis. In the present observations, DCO$^+$ 2--1 
seen in absorption is blended with an
emission line from CH$_3$OCHO (21(3,18)--21(2,19)) at 144.070 GHz emitted
at the velocity of the HII region at 55 km~s$^{-1}$. The 2--1 and 3--2 transitions are both, to our
knowledge, the first detections of the DCO$^+$ species seen in
absorption. We were able to fit both narrow absorption lines to obtain constraints
on the kinetic temperature and p-H$_2$ density using the collision rates
provided by the BASECOL database\footnote{http://basecol.obspm.fr/}, 
computed down to 5 K. For the absorption component, the density must be
$\le$ 10$^5$ cm$^{-3}$ (so that the 2--1 transition stays in absorption)
and the kinetic temperature must be $\le$ 10 K (or both transitions will
appear in emission). In order to reproduce both transitions (2--1 and
3--2) with a fixed temperature of 10 K we need a density of 2.5~10$^{4}$ cm$^{-3}$ (see Table
\ref{radtrans}). The ortho-para ratio of molecular hydrogen is a
  significant unknown; however, chemical studies for the cold and dense
  regions of the interstellar medium have shown a low ortho-para ratio
  \citep{dislaire2012} compatible with deuterium enrichment
  \citep{pagani2009}. The computed p-H$_2$ density is likely close to the
  H$_2$ density in the core study. From our DCO$^+$ observations, we can
conclude that the absorption traces a dense cloud,
which must be cold enough to become deuterated, along the line of sight. If DCO$^+$ had been detected with no background continuum, its line
  intensity for the 2--1 transition would have been 0.2 K. However, at such a
  large distance, such a small source would be heavily diluted in a single-dish telescope beam and would need observations with a mm
  interferometer. Absorption studies are best suited, and are not affected
  by beam dilution, as long as the absorbing layer is larger than the
  continuum.

For the H$^{13}$CO$^+$ transition we use the collision coefficients (down
to 10 K) provided by the LAMDA database \citep{flower1999}. With only
  one transition, we fixed the temperature and density as estimated from
  DCO$^+$ to estimate the column density. Using a $^{12}$C/$^{13}$C ratio
of 68 \citep{milam2005}, the resulting DCO$^+$/HCO$^+$ ratio is about 0.02.

DNC has also been detected in absorption against the CH$_3$OCHO transition
(19(3,17) --18(2,16)) at 228.894 GHz emitted from the HII region. We used
the collision coefficients from \citet{dumouchel2011} available in the
  BASECOL database, and fixed the kinetic temperature at 10 K  and
the p-H$_2$ density at 2.5~10$^{4}$ cm$^{-3}$. We used the same strategy
for HN$^{13}$C which has been detected in absorption at the same
velocity. The DNC/HN$^{13}$C ratio is surprisingly high, $\sim$ 9.2,
  resulting in a DNC/HCN ratio of 0.14 using the same assumptions as
for DCO$^+$. However, \citet{roueff2015} showed that the $^{12}$C/$^{13}$C
ratio in nitriles may be different from the isotopic ratio. Hence, the
DNC/HNC ratio may be somewhat lower.

The o-NH$_2$D species is the only deuterated species that we detected
  in emission at $\sim$ 69 km~s$^{-1}$. However, its velocity is slightly different
from that of  the other species, which makes it difficult to explain in
terms of the geometry of the source. Assuming no beam dilution
  in our IRAM 30m beam we derived the column density within the [5-10] K
  range (see Table \ref{radtrans}). However, our NOEMA observations of the
  same ortho-NH$_2$D transition, resulted in the detection of two adjacent
  cores. Using the same temperature range we also derive the column
  densities for both Ellipse A and Ellipse B detections (see Table \ref{pdb}).

The excitation of the two p-H$_2$CO transitions detected in absorption at
about 69 km~s$^{-1}$ with our IRAM observations is compatible with the
physical conditions derived from DCO$^+$. The collisional rates used for
p-H$_2$CO are taken from \citet{wiesenfeld2013}, computed for
  temperatures as low as 10 K. The physical conditions derived from
  the  detected transitions are slightly different from the values
  derived for CN by \citet{mookerjea2014} with T$_K$ $\sim$ 6 K and
  n(H$_2$) $\sim$ 10$^5$ cm$^{-3}$. However, owing to the physical complexity
  of the W51 molecular cloud complex, it is possible that we detected many
  components with different molecular probes (D-species versus CN). Also,
  with the contamination by the W51 molecular cloud complex, the CN line
  profiles were difficult to analyze, leading to a 50\% uncertainty on the
  kinetic temperature and H$_2$ density.

\begin{table} 
  \tiny
  \caption{Column densities and abundances derived from LTE or non-LTE radiative transfer.\label{radtrans}}
  \begin{tabular}{cccccc}
    Species                 &  V$_{LSR}$      &    T$_{ex}$            &  T$_{k}$ & n(H$_2$)             &            N             \\
    &    (km.s$^{-1}$  &        (K)                  &    (K)       &  (cm$^{-3}$)          &    (cm$^{-2}$)                  \\
    \hline
    \hline
    DCO$^+$              &   69.58   &        3.5\tablefootmark{a}                  &  10          &   2.5~10$^{4}$     &   5~10$^{11}$  \\
    H$^{13}$CO$^+$  &   69.64   &        3.5                   &  10          &  2.5~10$^{4}$      &   4~10$^{11}$    \\
    DNC                      &   69.52   &   3.0\tablefootmark{b}                  &       10         &    2.5~10$^{4}$                         &  1.2~10$^{12}$   \\
    HN$^{13}$C          &   69.70   &         3.1       &  10   & 2.5~10$^{4}$  &  1.3~10$^{11}$   \\
    o-NH$_2$D           &   68.91   &    5--10                &                &                              &  0.4--1.8~10$^{12}$\tablefootmark{c}  \\
    &   68.90   &    5--10                &                &                              &  0.2--1.1~10$^{13}$\tablefootmark{d}  \\
    &   68.54   &    5--10                &                &                              &  1.3--6.9~10$^{12}$\tablefootmark{e}  \\
    p-H$_2$CO           &   69.39   &      3.5\tablefootmark{a}                         &  10         & 2.5~10$^{4}$        &  5~10$^{12}$   \\

    \hline
  \end{tabular}
  \tablefoot{\tablefoottext{a}{For both transitions}; \tablefoottext{b}{at higher T$_{ex}$ the line appears in emission}; \tablefoottext{c}{using the IRAM 30m with no dilution}; \tablefoottext{d}{using the NOEMA observations for Ellipse A}; \tablefoottext{e}{using the NOEMA observations for Ellipse B}.}
\end{table}

\subsection{Comparison with a typical dark cloud: TMC-1}

The resulting temperature (10 K) and H$_2$ density (2.5~10$^{4}$ cm$^{-3}$)
for the $\sim$ 69 km~s$^{-1}$ narrow component is compatible with dark
cloud conditions. Strong deuteration of molecules and ions (e.g., DNC and
DCO$^+$) has been recognized as a landmark of these regions that are
the birthplaces of low-mass stars. The molecule H$_2$ (respectively HD) is the main
reservoir of hydrogen (respectively deuterium) in dark clouds. Gas-phase
deuteration at low temperature is driven by the molecular ions H$_2$D$^+$
and D$_2$H$^+$ \citep{millar1989,vastel2004}, which play a major role in $\le$
20 K gas. In prestellar cores, the freezing of CO on grains implies a
  reduction of the production rate of HCO$^+$ since the main production
  route is the protonation of CO. However, for DCO$^+$, the significant
  increase in the abundance of H$_2$D$^+$, D$_2$H$^+$, and D$_3$$^+$
  counterbalances the decrease in CO abundance. The DCO$^+$ abundance
  remains more or less constant, as shown by Vastel et al (2006) in the
  case of L1544 and further discussed by \citet{pagani2011}.  In the
  external layers of prestellar cores (radius between 5000 and 10000 AU),
  where the density approaches 10$^4$ cm$^{-3}$ and the kinetic temperature
  remains close to 10 K, the depletion of CO is less extreme than in the
  central region. These dark cloud
  conditions are well suited for the production of DCO$^+$.\\


We can now compare W51-core with TMC-1, a prototypical dark cloud located
in the Taurus molecular cloud at a distance of 140 pc
\citep{elias1978,onishi2002}. The most massive molecular cloud in Taurus is
the Heiles Cloud 2 (HCL 2; \citet{heiles1968}; \citet{onishi1996};
\citet{toth2004}). It hosts several cold clumps first recognized in
extinction maps and recently detected by the \citet{planck2015}.  TMC-1 is
a long and dark filament, running approximately SE to NW with a temperature of
about 10 K and a mean column density of $\sim$ 2~10$^{22}$ cm$^{-2}$
\citep{feher2016}. TMC-1 shows a remarkable variation in the chemical
abundances; there are carbon chains and cyanopolyynes in the SE part, notably in
the cyanopolyyne peak TMC(CP), and ammonia peaks in the NW regain, which
shows a more evolved chemistry. Deuteration is lower in the CP region
\citep{turner2001} than in the ammonia peak \citep{tine2000} where
DCO$^+$/(H$^{13}$CO$^+$ $\times$ 68) $\sim$ 0.011, which is similar to the value
found for W51-core.
However, the DNC/HNC ratio found by \citet{hirota2001} is a factor of 5
lower than the value we found in W51-core (0.028 vs. 0.15).

\subsection{Increased deuterium fractionation for HNC}

Hirota et al. (2001) observed 29 nearby dark cloud cores and measured a
large variation in the DNC/HN$^{13}$C ratio between 0.5 and 7.3
(DNC/HNC=[0.01--0.11]). The lowest ratios correspond to the
carbon-chain producing regions where CCS and cyanopolyynes are abundant. In
addition, they found a gradient in TMC-1 where the DNC/HN$^{13}$C ratio
in the northwest part of the TMC-1 ridge including the NH$_3$ peak is found
to be higher than that in the  southeast part around the cyanopolyyne
peak. These authors found no apparent relation between the
DNC/HN$^{13}$C ratio and the H$_2$ density and kinetic temperature, but
concluded that the deuterium fractionation of DNC may be related to the
chemical evolution of dark cloud cores.

These values are higher than those predicted by pure gas-phase chemical
models for cold dark cloud cores \citep{roueff2015}. The effect of the
depletion of molecules onto grain surfaces is also invoked to explain the
observed ratio. In the L1544 prestellar core, a large CO depletion was
reported by \citet{caselli2002}, leading to a high deuterium
fractionation. The DCO$^+$/HCO$^+$ ratio is much enhanced toward the center
of the core although the DNC/HNC ratio is constant throughout the cloud
\citep{hirota2003}, leading to the conclusion that DNC/HNC is not very
sensitive to the depletion factor unlike DCO$^+$/HCO$^+$ and
N$_2$D$^+$/N$_2$H$^+$. The variation in the DNC/HN$^{13}$C ratios from core
to core is much larger than that within each core. This variation might be
related to the timescale of the deuterium fractionation and the observed
variation from one core to another might be the combination of the three
competing, time-dependent processes: gas-phase deuterium fractionation,
depletion of molecules onto grain surfaces, and dynamical evolution of a
core (see, e.g., Aikawa et al. 2001). The DNC/HNC ratio becomes higher for
more chemically evolved cores.

More recently \citet{sakai2012} has observed 18 massive clumps including infrared dark clouds (IRDCs) and high-mass protostellar objects (HMPOs). The
averaged DNC/HNC ratio in HMPOs is lower than in IRDCs, with higher kinetic temperature for HMPOs. However, 
some of the IRDCs with a Spitzer 24 $\mu$m source show a lower DNC/HNC ratio than that of HMPOs, although the 
temperature is lower. This suggests that
DNC/HNC  does not depend on the kinetic temperature only. However, chemical models
suggest that the DNC/HNC ratio  decreases after the birth of
protostars, depending on the physical conditions and history in their starless phase. More recent ALMA observations of IRDC G34.43+00.24 show a
DNC/HNC of 0.06, which is significantly larger than the value of 0.003
determined from single-dish observations \citep{sakai2015}.  They suggest
that single-dish observations may trace low-density envelopes while ALMA
traces higher density and highly deuterated regions.  Chemical model
calculations from \citep{sakai2015}, based on the chemical reaction
  network of K. Furuya et al. (in preparation), show that at a given
time, the DNC/HNC abundance ratio is significantly higher in the dense core 
than in its immediate environment.

Additionally, \citet{roueff2015} have shown that a higher ortho-para ratio for
H$_2$ (typically in less dense regions) leads to a higher DNC/HNC
ratio, demonstrating the complexity of the interplay between the chemistries of the species D, $^{13}$C,
and $^{15}$N.

\subsection{Distance of the W51-core}

Infrared
dark clouds have filamentary structures that are seen in absorption in the mid-infrared
against the bright Galactic background. They are complexes of cold ($\le$
25 K) and dense ($\ge$ 10$^{3-4}$ cm$^{-3}$) molecular gas within which
cold cores of typical size 0.1 pc have recently been detected
\citep[e.g.,][]{ragan2015}. We possibly detected an IRDC as the filamentary
structure in front of W51 where prestellar cores and their
surrounding clump material are interacting. The absorption feature at
68\,\kms\ has a velocity forbidden by the circular galactic rotation model
\citep{schmidt1965} and is thought to be due to the gas streaming along
the spiral arm in the density-wave
theory of galactic rotation.

According to the density-wave theory, the HV streaming gas is located near
the subcentral point, but extends over 1.5\,kpc along the line of sight
\citep{koo1997}. All existing observations indicate that the HV stream is
in front of the HII regions.  Given that the stream shows evidence of
interaction with the W51 main cloud triggering star formation it has to be
physically interacting with the main cloud \citep{kang2010}.

Thus, W51-core which we detect in the HV stream is likely located at a
distance of not less than 4\,kpc. This makes it the most distant dark cloud
detected so far in the Galaxy.  We can use H$^{13}$CO$^+$ as a probe of the
total H$_2$ column density, with an HCO$^+$ abundance of 3~10$^{-9}$ and a
$^{12}$C/$^{13}$C ratio of 68, leading to a value of N(H$_2$) = 9~10$^{21}$
cm$^{-2}$, compatible with dark cloud conditions.
 From our NOEMA observations, two nearby fragments (separated by less
  than 10$^{\prime\prime}$, corresponding to about 0.19--0.24 pc at a
  distance of 4-5 kpc) have been identified. The fragments are compact,
   less than 0.16--0.19 pc in size (at a distance of 4--5 kpc). Recently,
  many high spatial resolution studies of IRDCs have led to the conclusion that
  these sources are highly fragmented, with regularly spaced cores; for example,  the 
  mean separation between the cores is  0.18 pc   in IRDC G035.39--00.33 \citep{henshaw2016}, 0.18 pc  in G011.11--0.12 \citep{ragan2015}, and
  (0.40$\pm$0.18) pc  in IRDC 18223 \citep{beuther2015}. These values are
  compatible with our findings. We note that a detailed study of the
  fragmentation of the filament leading to the formation of the W51-core is
  limited because of the continuum and line contamination from the W51
  compact HII region. 

\section{Conclusions}

 Combined Herschel/HIFI and IRAM 30m observations led to the 
 detection of a distant dark cloud (4--5 kpc) along the line of sight of a
  compact HII region W51. This cloud, named W51-core, presents a fragmented
  structure that was determined from our IRAM/NOEMA observations of the NH$_2$D
  transition seen in emission and has  two cores (less than 0.16--0.19
  pc in size) separated by 0.19--0.24 pc. These high spatial observations of
  W51-core are consistent with a filamentary structure in front of W51,
  fragmenting while interacting with the giant molecular cloud, leading to
  the formation of many cold cores trailing along the filament. Our
  IRAM 30m observations led to the detection of W51-core in absorption
  against the strong continuum emitted by W51. Although not affected by
  beam dilution, absorption studies cannot trace the spatial extent of the
  source. From the IRAM 30m  observations we derived the physical properties
  of W51-core with a density of $\sim$ 2.5~10$^4$ cm$^{-3}$ and a
  temperature of about 10 K. It presents a high deuterium fractionation of
  HCO$^+$ (0.02) and HNC (0.14) with the first detection of DCO$^+$ in
  absorption.  Because of  its fortuitous location along the line of sight
  with the continuum-bright W51e2 compact HII region, W51-core has not been
  detected in the continuum images. However, it makes it an appropriate
  source for absorption studies such as the search of H$_2$D$^+$,
  D$_2$H$^+$, and D$_3^+$.

\begin{acknowledgements}
  C.V. is grateful for the help of the IRAM staff at Granada during the
  data acquisitions, and also for their dedication to the
  telescope. B.M. acknowledges the support received from University of
  Toulouse for her stay in Toulouse, which enabled the completion of this
  work. This work was supported by the CNRS program "Physique et Chimie du
  Milieu Interstellaire" (PCMI).
\end{acknowledgements}

%
%

{}

\end{document}